\documentclass[final,5p,times,twocolumn]{elsarticle}
\usepackage{hyperref}
\usepackage{amssymb}
\usepackage{amsmath}

\usepackage{url}

\journal{Nuclear Inst. and Methods in Physics Research, A}

\usepackage{xcolor}
\usepackage{soul}

\begin{document}

\begin{frontmatter}

\title{Investigation of radiation emitted by sub GeV electrons in oriented scintillator
crystals}

\author[inst2]{L. Bandiera \corref{corr2}}
\author[inst2]{R. Camattari}
\author[inst2]{N. Canale}
\author[inst3, inst4]{D. De Salvador}
\author[inst1, inst2]{V. Guidi}
\author[inst6]{P. Klag}
\author[inst6]{W. Lauth}
\author[inst2]{A. Mazzolari}
\author[inst1,inst2]{R. Negrello \corref{corr}}
\author[inst2]{G. Paternò}
\author[inst2]{M. Romagnoni}
\author[inst4]{F. Sgarbossa}
\author[inst1,inst2]{M. Soldani}
\author[inst2]{A. Sytov}
\author[inst5]{V. V. Tikhomirov}

\cortext[corr]{Corresponding author; \textit{Email address:} \texttt{riccardo.negrello@unife.it}}

\cortext[corr2]{Corresponding author; \textit{Email address:} \texttt{bandiera@fe.infn.it}}

\affiliation[inst1]{organization={Department of Physics and Earth Science, University of Ferrara},
            addressline={Via Saragat 1}, 
            city={Ferrara},
            postcode={44122}, 
            state={Italy}}

\affiliation[inst2]{organization={INFN Section of Ferrara},
            addressline={Via Saragat 1}, 
            city={Ferrara},
            postcode={44122}, 
            state={Italy}}
            
\affiliation[inst3]{organization={Department of Physics, University of Padua},
            addressline={Via Marzolo 8}, 
            city={Padua},
            postcode={35131}, 
            state={Italy}}

\affiliation[inst4]{organization={INFN Laboratori Nazionali di Legnaro},
            addressline={Viale dell'Università 2}, 
            city={Legnaro},
            postcode={35020}, 
            state={Italy}}
            
\affiliation[inst5]{organization={Institute for Nuclear Problems, Belarusian State University},
            addressline={Bobruiskaya 11}, 
            city={Minsk},
            postcode={220030}, 
            state={Belarus}}

\affiliation[inst6]{organization={Institut für Kernphysik der Universität Mainz},
            city={Mainz},
            postcode={55099}, 
            state={Germany}}

\begin{abstract}

The research investigates coherent interactions between sub-GeV electrons and oriented scintillator crystals, leading to enhanced electromagnetic (EM) radiation. Experiments at Mainz Mikrotron (MAMI) involved PWO, BGO, and CsI crystals oriented along $\langle100\rangle$, $\langle111\rangle$, and $\langle100\rangle$ axes. Enhanced radiation emission was observed when the beam aligned with crystal axes, especially in BGO and CsI for the first time. These findings are crucial for innovative detectors using oriented crystal scintillators, amplifying EM processes along specific crystallographic directions. Potential applications include ultra-compact, highly sensitive electromagnetic calorimeters for high-energy physics and astroparticles, as well as high-performance gamma detectors for nuclear physics and medical imaging.
\end{abstract}


\end{frontmatter}

\section{Introduction}
\label{sec:sample1}

In particle physics, inorganic scintillator crystals are commonly used as gamma, charged particle detectors, as well as electromagnetic (EM) calorimeters. As an example of a state-of-the-art crystal EM calorimeter, the ECAL for the Compact Muon Solenoid (CMS) experiment at CERN, one of the two experiments that discovered the Higgs boson in 2012 at the Large Hadron Collider (LHC), consists of 80000 crystals of lead tungstate ($PbWO_4$, abbreviated PWO) \cite{CMSExperiment}. Bismuth Germanate crystals ($Bi_4Ge_3O_{12}$, abbreviated BGO), are used in nuclear medicine for tumour diagnosis through Positron Emission Tomography or in the Dark Matter Particle Explorer (DAMPE) \cite{dampe}. In many space-borne experiments, such as the Fermi Gamma-ray Space Telescope \cite{Atw}, a Cesium Iodide (CsI) crystal e.m. calorimeter is one of the key sub-detectors, designed to perform measurements over a wide range in energy from a few GeV up to several TeV.  
The interaction between charged particles and crystalline medium varies significantly based on the angle between particle trajectories and crystal orientation\cite{TerMikaelian,BaierKatkov}. Random orientation results in classical interaction, producing a bremsstrahlung spectrum. However, when particles align closely with a crystal plane or axis, they experience periodic potential peaks. Negatively charged particles bind to atomic nuclei, while positively charged ones oscillate between nearby axes, leading to channeling effects and specific radiation emission .
The so-called Channeling Radiation (CR) typically displays a dipole character at sub-GeV energies\cite{Lin, Kumakhov197617, BARYSHEVSKY198061}, increasing in intensity to the low-energy region of the bremsstrahlung spectrum\cite{EPJC2021,Ban3, Ban}. In contrast, a synchrotron-like spectrum is formed at a few-GeV or higher, with a further increase in the energy of the enhanced spectral components\cite{WIENANDS201711, PhysRevLett.121.021603, EPJC2022, Gui}.
The angular acceptance of the channeling effect, the Lindhard angle, can be calculated using the following formula:
\begin{equation}
\label{thetaC}
    \theta_C = \sqrt{\dfrac{2U_0}{pv}}
\end{equation}
where $U_0$ is the depth of the averaged atomic potential calculated for the specific axis/plane, while $pv$ for relativistic particles is basically its energy. If a PWO crystal is considered, along $\langle100\rangle$ axis, the value of $U_0$ is approximately 470 eV, therefore, if incident particle energy is about 1 GeV, the value for the Lindhard angle is about 1 mrad.
At higher particle energies, the electromagnetic field to which charged particles are subjected is Lorentz boosted by a factor $\gamma=\frac{\epsilon}{mc^2}$, where $m$ is the mass of the particle. The so-called Strong Field regime is achieved when $\frac{\gamma E}{E_0}>1$, where $E$ is the electric field along the axis in the laboratory frame, and $E_0\approx 1.32\times10^{18}\, \mathrm{V/m}$ is the Schwinger QED critical field, above which nonlinear effects occur in vacuum. In comparison to the incoherent condition, the interactions with the crystalline structure in the Strong Field regime highlight the emission of quantum synchrotron radiation, which exhibits higher intensity and a boost of the hard-photon component.
The study was conducted at energies that prevent the Strong Field effect, to verify and quantify the increase in radiation produced, even at low energies, for these high-Z scintillator crystals when specific crystal axes are aligned with the incident beam compared with the random (amorphous) orientation case.
Here, experimental results of enhancement of radiation production due to coherent effects in PWO, CsI, and BGO scintillator crystals, for the [100], [111], and [100] axes respectively, are shown. The data were collected at the Mainz Microtron (MAMI) facility, by aligning the crystal axes with a electron beam tuned to 855 MeV.

\section{Pre-characterization}

The scintillator crystals were tested through hard X-ray at the ID11 and BM05 beamlines at the European Synchrotron Radiation Facility (ESRF) of Grenoble (France) to measure their lattice quality. Indeed, the crystallographic perfection is mandatory to exploit axial effect for the enhancement of electron beam interaction in an oriented scintillator crystal. The performed characterizations are described here below.

\subsection{Hard X-ray diffraction}
\label{sec:anxdiff}
\begin{figure}[ht]
\begin{center}
    \begin{tabular}{c}
        \includegraphics[width=0.5\linewidth]{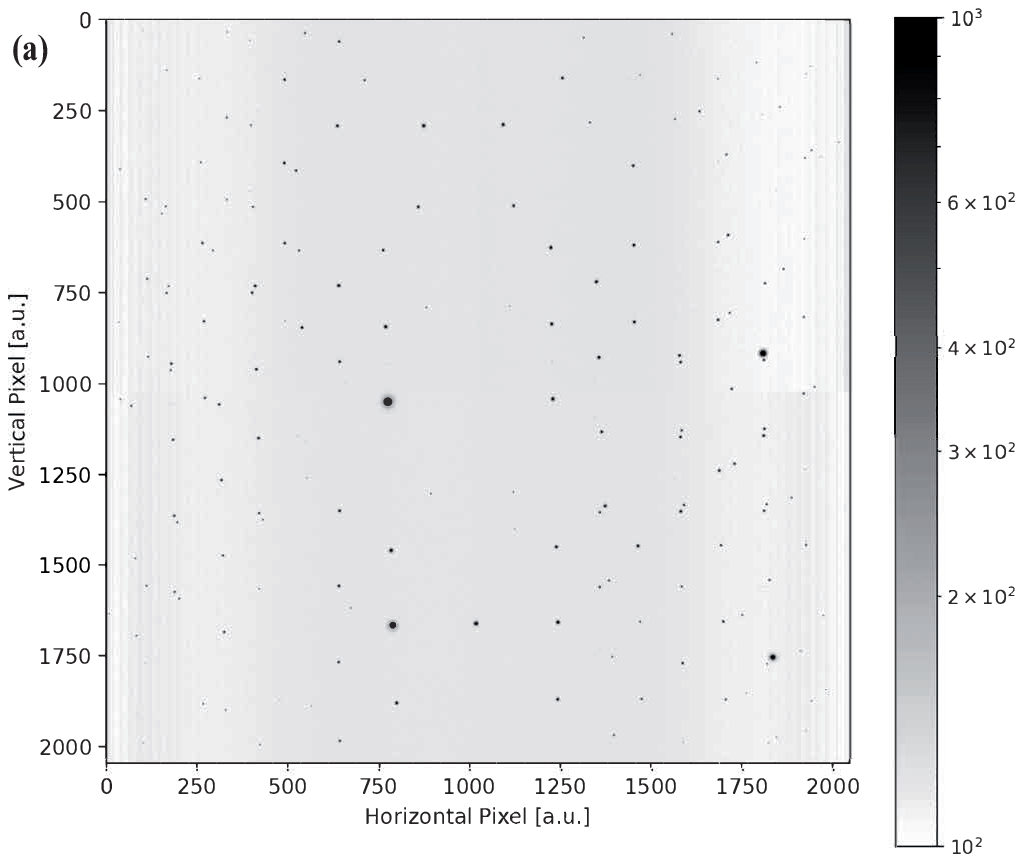} 
        \includegraphics[width=0.5\linewidth]{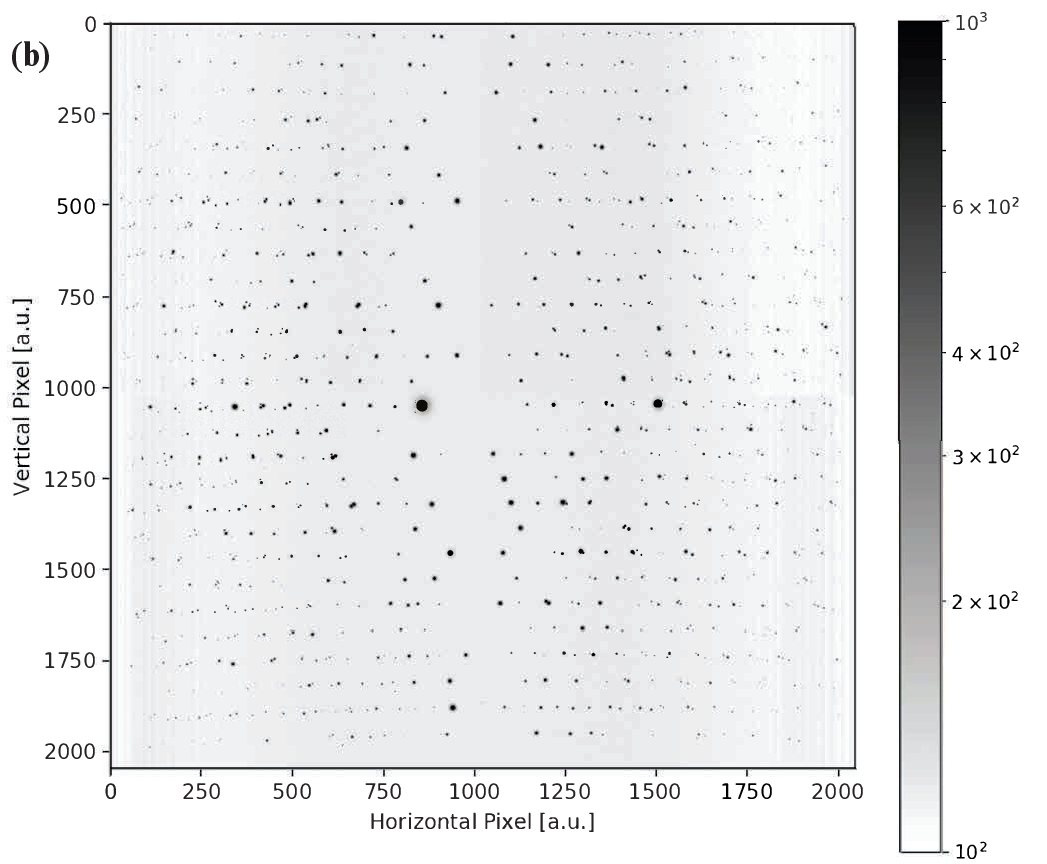} \\
        \includegraphics[width=0.5\linewidth]{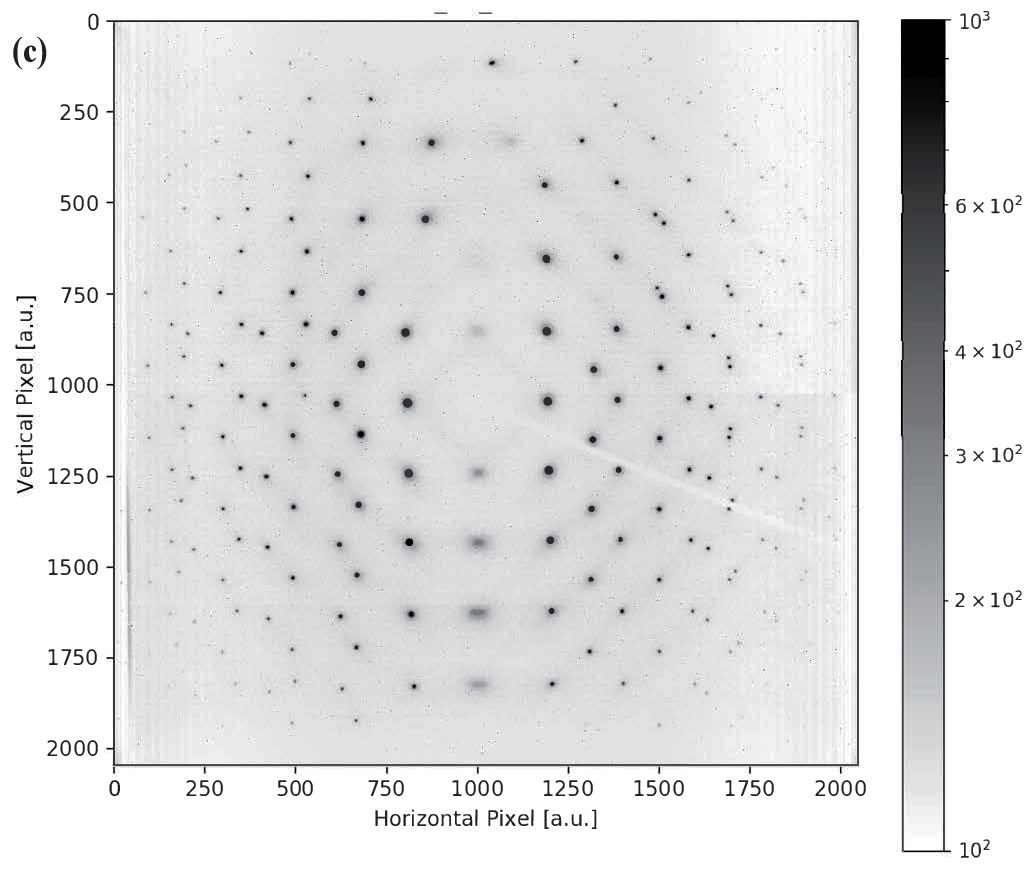} 
    \end{tabular}
    \caption{Photographic plate measurements. (a) PWO; (b) BGO; (c) CsI. The crystallographic structure is visible.} \label{fig:plate}
    \end{center}
\end{figure}

A highly monochromatic and quasi-parallel X-ray beam 50$\times$50 $\mu$m wide was tuned to 140 keV energy. The monochromaticity was of the order of $\Delta$E/E = 10$^{-3}$. The beam traversed the crystal thickness. The first measurement consisted in a photographic plate that was impressed with the X-rays that were diffracted by the sample while it was rotated about a vertical axis, perpendicular to the X-ray beam. This measure was used for a preliminary check of the crystallographic orientation and quality of the samples. The results are shown in Fig. \ref{fig:plate}.

\begin{figure}
\begin{center}
        \includegraphics[width=0.48\linewidth]{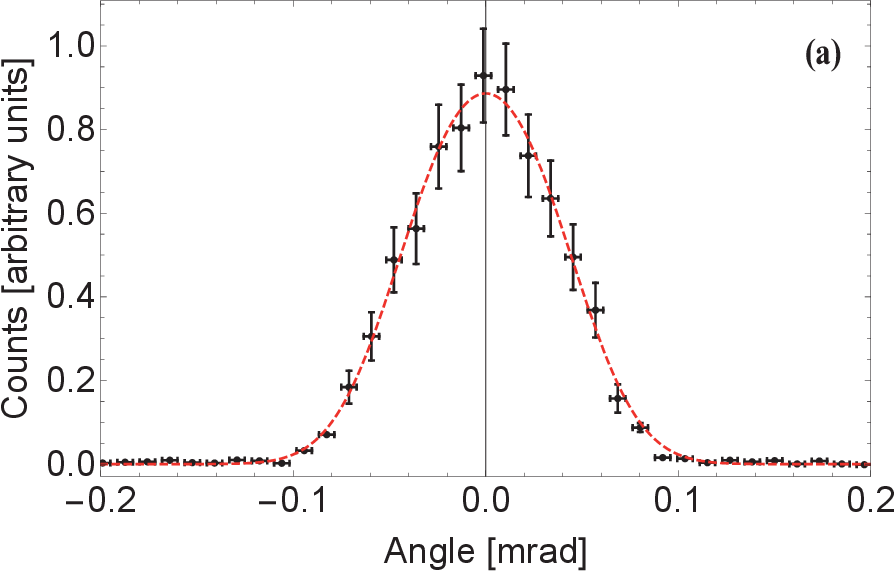} 
        \includegraphics[width=0.48\linewidth]{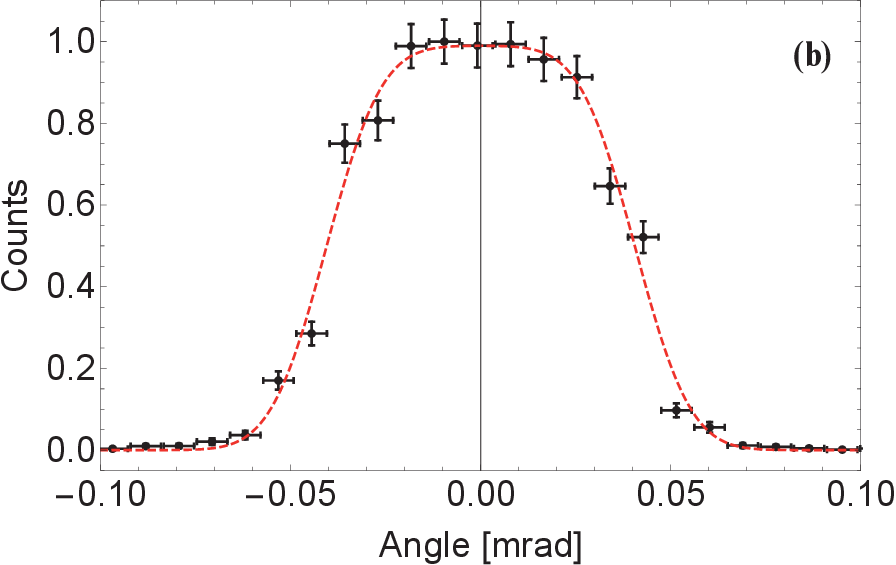} \\
        \includegraphics[width=0.48\linewidth]{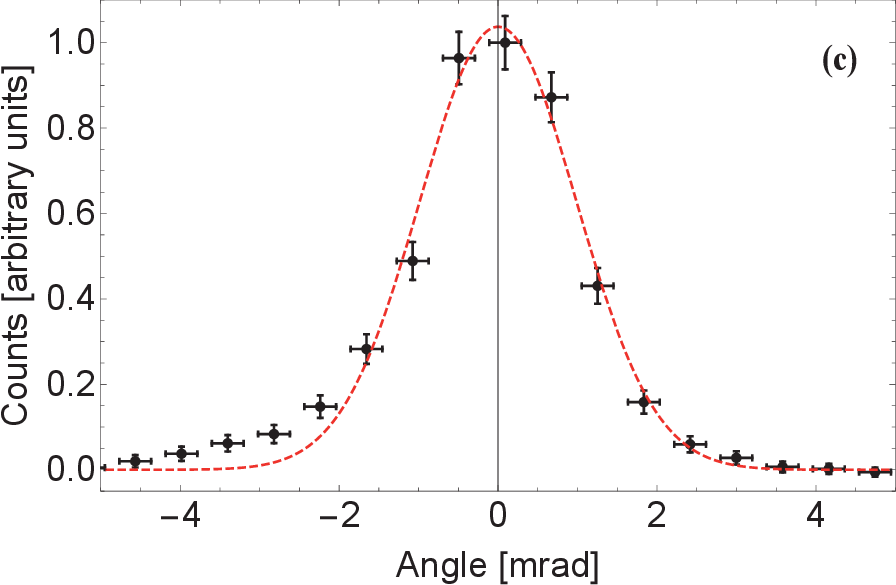} \\
    \caption{RC measurements. (a) PWO; (b) BGO; (c) CsI. The dashed red lines are the fitted function to find the mosaicity of the samples.} \label{fig:RC}
    \end{center}
\end{figure}

Subsequently, the nearest diffraction points were analyzed with the same beam via Rocking Curves (RC), \textit{i.e.} by recording the diffracted beam intensity while the crystal was being rotated around the position where Bragg condition was satisfied. The diffraction RC was recorded by a photo-diode as a function of the beam glancing angle. The samples were set far enough from the detectors to allow sufficient separation of diffracted and transmitted beams. The results are shown in Fig. \ref{fig:RC}. As can be seen, the values of Full Width at Half Maximum (FWHM) of the RCs, which are a good indicator of the mosaicity of the samples and then of their crystalline quality, is very low, namely 0.088, 0.081, and 2.3 mrad for the PWO, the BGO, and the CsI sample, respectively. The FWHM values have been found by fitting the RCs (red dashed lines in the figures).

\subsection{X-ray topography}

The crystalline quality of the scintillator samples were also measured through monochromatic X-ray topography at the BM05 beamline of ESRF. In particular, the samples were analyzed through Rocking Curve Imaging in reflection mode (Bragg geometry). 

\begin{figure}[h!]
\begin{center}
  \includegraphics[width=0.4\linewidth]{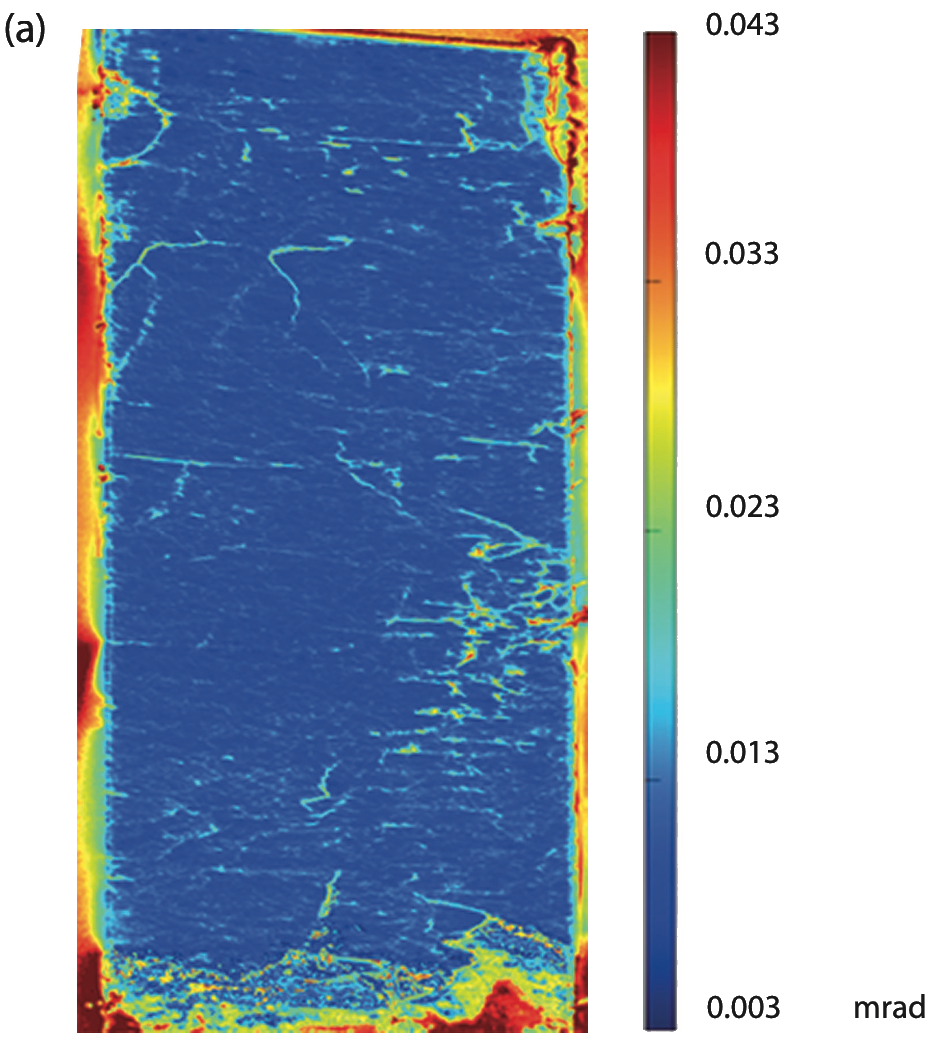}
  \hfill
    \includegraphics[width=0.4\linewidth]{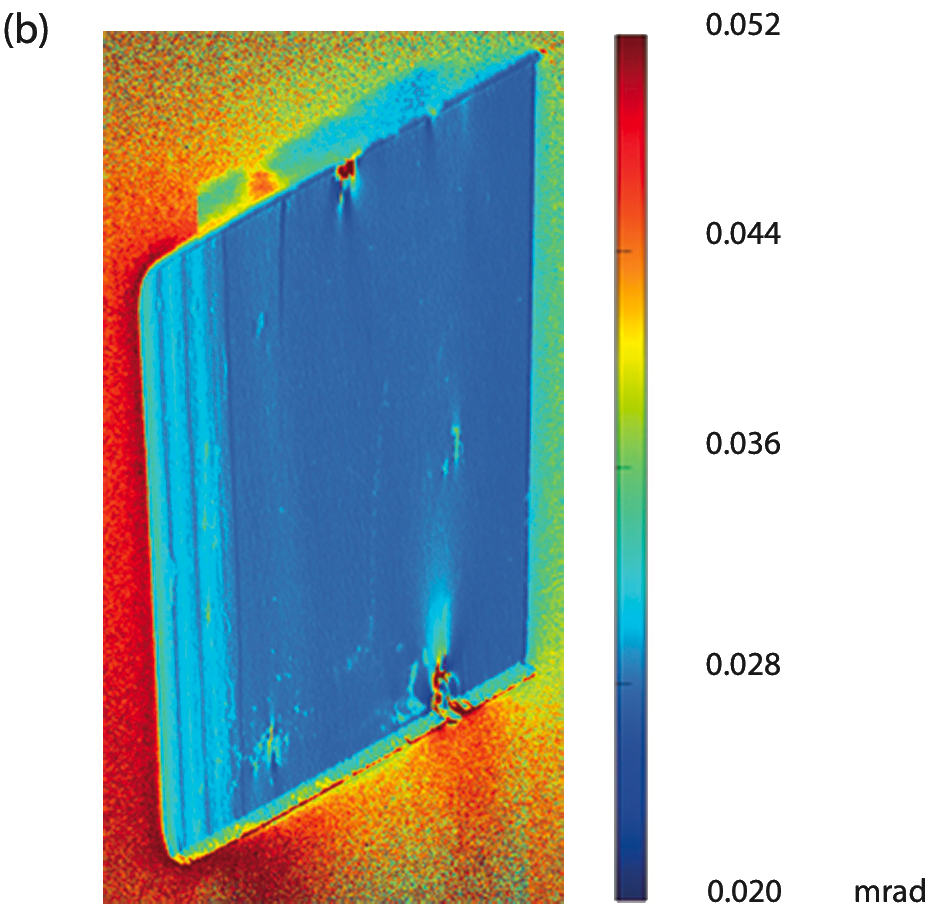}
  \caption{X-ray topography. (a) PWO; (b) BGO.}\label{fig:BM05}
\end{center}
\end{figure}

The beam was set to 20 keV with a monochromaticity of the order of $\Delta$E/E = 10$^{-4}$, and was 10$\times$10 mm wide. The sample had to be rotated around the position where Bragg condition was satisfied. The detector was a FReLoN camera with pixel size of 5$\times$5 $\mu$m. The image size was 100$\times$100 mm, however in case of reflection the image size corresponds to 10 mm in the horizontal direction and 10mm/sin(Bragg angle) in the vertical direction. Thus, the images turned out to be compressed along the vertical direction. The sample surfaces were analyzed by stitching all the measurements.
The result of the measurement consisted in a map of the sample surface. In particular, for every point of the surface, a RC was recorded. Fig. \ref{fig:BM05} consists in a map of all the FWHMs of the corresponding RCs. The sample surface turned out to be quite homogeneous, with a mosaicity compatible with that found during RC measurements. Some superficial defect is visible in red, where the mosaicity is increased. Unfortunately, the measurement of X-ray topography could not be performed on the CsI sample, probably because its hygroscopic nature ruined its surface enough to not allow investigation of its surface.

\begin{figure}
\begin{center}
  \includegraphics[width=0.5\textwidth]{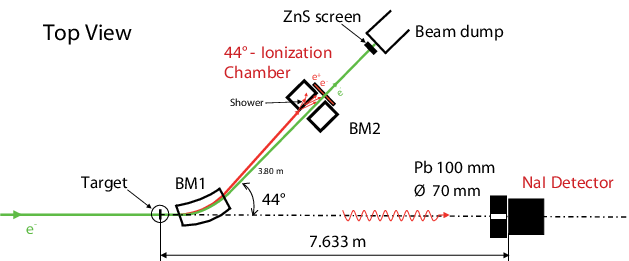}
  \caption{Experimental setup, top view. Photon spectra are detected with a NaI detector. The detector is shielded by a 100 mm thick lead wall with a 70 mm opening for the photons. The 44$^\circ$-ionization chamber, filled with air at standard pressure, is employed to detect crystal alignment. Downstream the crystal target, the beam is deflected horizontally by the 44$^\circ$ bending magnet BM1 and vertically by a 7.2$^\circ$ bending magnet BM2. Showers are produced by the electrons that leaved the nominal direction due to scattering or energy loss in the target (in red). Just in front of the beam dump, the beam spot can be monitored with a ZnS luminescent screen which is viewed by a CCD camera.}\label{fig:setup}
\end{center}
\end{figure}

\section{Experimental method}

After the pre-characterization that attested the high crystallographic quality of the scintillator samples, an experiment to verify the increase of interaction between an electron beam and the samples in case of axial alignment was performed at the Mainz Microtron (MAMI) at Mainz, Germany. The experimental setup is shown in Fig. \ref{fig:setup} (see \cite{MAMISetup,SYTOV2017}). The low emittance electron beam at MAMI was tuned to 855 MeV, resulting in a pencil beam with dimensions of 10$\times$10 $\mu m^2$ and small angular divergence, boasting an emittance of just 1 $\mu rad*mm$. The crystals were mounted on a goniometer with which rotations around two axis could be accomplished. The electron beam traversed the scintillator sample through its thickness, along specific axis (see Tab. \ref{tab:sample_parameters}), thus generating channeling radiation. Then, the photons emitted by the electrons inside the sample were separated by the charged beam through a bending magnet (BM1) and after 7.633 m arrived at a 25.4$\times$25.4 cm$^2$ NaI scintillator detector. An aperture of 70-mm diameter in the lead shield that surrounded the detector permitted the collection of a portion of the emitted photons, resulting in a collimator aperture of 4.63 mrad,\textit{ i.e.}, equal to $\sim$7.8 times the 1/$\gamma$ angle, to collect most of the spectrum of the emitted photons.

The alignment of the electron beam with the axes of the crystalline target was accomplished through a 44$^\circ$-ionization chamber behind the bending magnet BM2. This chamber is sensitive to the charged particles of e.m. showers which are produced by the electrons that leaved the nominal beam direction due to scattering or after emission of multi-MeV photons in the target crystal. Thus, when the crystal axes were aligned with the electron beam, a clear signal was visible via the ionization chamber. Finally, beam-off background has been subtracted from the obtained spectra. In order to measure the radiation spectrum, a calibration of the NaI detector was performed by using the natural radioactive isotopes $^{40}$K (1.461 MeV) and $^{208}$Tl-$^{228}$Th (2.6146 MeV). The measured samples were a PWO, a BGO, and a CsI scintillator crystals. All the specifications of the samples are listed in Table \ref{tab:sample_parameters}.
\begin{table}
\begin{center}
\begin{tabular}{c|ccc}
    Sample material & PWO & BGO & CsI \\
    \hline
    Lateral size (mm$^2$) & 8$\times$8 & 10$\times$10 & 15$\times$15 \\
    Thickness (mm) & 0.5 & 0.1 & 1 \\
    Main surface & (100) & (100) & (100) \\
    Axes & [100] & [111] & [100] \\
\end{tabular}
\caption{Sample parameters}
\end{center}
\label{tab:sample_parameters}
\end{table}


\section{Results and discussion}

\begin{figure}
\begin{center}
  \includegraphics[width=0.5\textwidth]{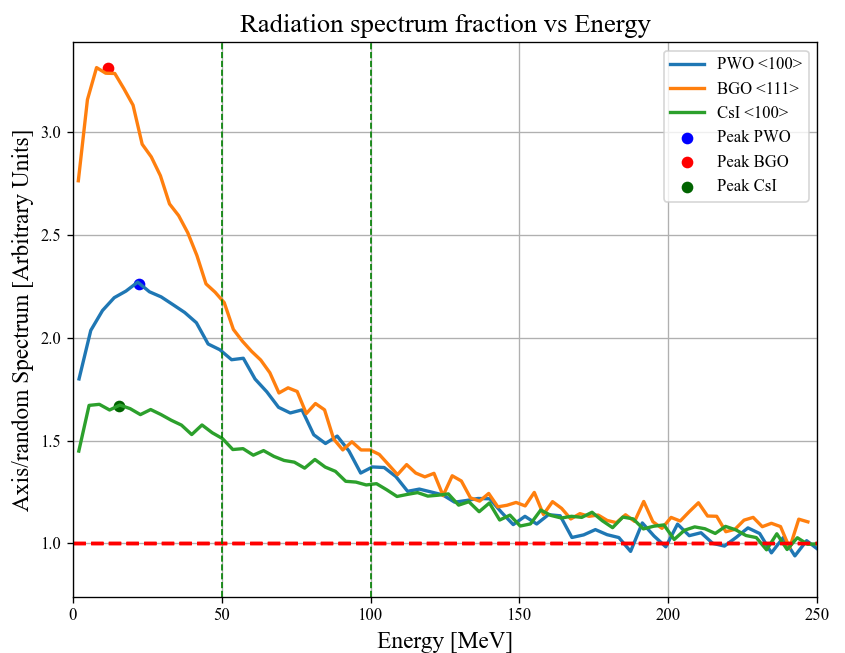}
    \caption{This figure presents experimental data collected at MAMI, depicting the ratios of radiation spectra obtained for each crystal. The spectra were produced by directing the beam along specific crystallographic directions, and the results are compared with the spectrum obtained when the crystal was randomly oriented to simulate an amorphous case. Enhancement peaks are denoted by colored dots, with green dashed vertical lines indicating the reference energies at 50 and 100 MeV. The amorphous spectrum line, normalized to 1, is represented by the red dashed line. }\label{fig:mami}
    \end{center}
\end{figure}

The experimental radiation spectrum resulting from crystal interactions is normalized against the amorphous case, revealing substantial photon increase (at 855 MeV), indicating enhanced electromagnetic processes in axially oriented crystals (Fig. \ref{fig:mami}). The BGO crystal along $\langle111\rangle$ exhibits the highest peak enhancement, followed by PWO and CsI. CsI's lower enhancement is attributed to high mosaicity, hindering coherent interactions and radiation enhancement (sec. \ref{sec:anxdiff}, Fig. \ref{fig:RC}). PWO and BGO, with lower mosaicity, exhibit significant coherent effects and higher enhancements.

Enhancements exceed 1 even for 50 MeV and 100 MeV photons (Table \ref{tab:enhancement_table}). Peak enhancement values and energies for PWO and BGO, in the table, were calculated via Gaussian fitting, errors from the covariance matrix. CsI, with unfeasible Gaussian fitting, employed the statistical bootstrap method. This approach yielded maximum values and associated errors, linking peak energy to bootstrap result and representing error through bin size. Bootstrap method was also used to compute 50 MeV and 100 MeV enhancement values, along with uncertainties.

\begin{table}
\begin{tabular}{c|cccc}
Sample & $E_{peak}$ \small{(MeV)} & $R_{peak}$ & 50 \small{MeV} & 100 \small{MeV} \\ \hline
PWO    & 22.2$\pm$0.8         & 2.26$\pm$0.02    & 1.94$\pm$0.18    & 1.37$\pm$0.18     \\
BGO    & 11.6$\pm$0.5         & 3.31$\pm$0.02    & 2.51$\pm$0.30    & 1.63$\pm$0.30     \\
CsI    & 15.3$\pm$0.9         & 1.67$\pm$0.01    & 1.58$\pm$0.09    & 1.35$\pm$0.09    
\end{tabular}
\caption{Table of enhancement values, $E_{peak}$ refers to the energy at the maximum of the enhancement, shown in the $R_{peak}$ column. Also the values of the enhancement are included for 50 and 100 MeV photons energy.}
\label{tab:enhancement_table}
\end{table}

\section{Conclusions}


In this study conducted at MAMI, we have made measurements regarding the enhancement of electromagnetic radiation emitted by sub-GeV electrons in axially oriented scintillator crystals. This research marks the first-ever measurements demonstrating the radiation enhancement resulting from coherent orientational effects in commonly used detectors in particle, nuclear, and medical physics, specifically BGO and CsI. Our investigation of the crystalline quality of the samples using HRXRD at the ESRF revealed significant findings.
Our study explores the possibility of utilizing these materials as gamma-radiators in crystal-based light sources. This promising avenue could revolutionize the field, offering enhanced capabilities and expanding the potential applications\cite{procChan2018}.
Our discoveries have far-reaching implications, in the development of gamma detectors with enhanced efficiency under specific directions due to the heightened pair production probability. These advancements have significant applications in nuclear and astrophysical domains, offering advantages such as reduced mass and enhanced angular selectivity. Furthermore, our research has paved the way for the creation of compact electromagnetic calorimeters for particle and astroparticle physics. In these crystalline detectors, bremsstrahlung for electrons and positrons (e$^{-}$/e$^{+}$) and pair production by photons are markedly amplified. Moreover, the electromagnetic shower length experiences a drastic reduction under particular crystal orientations. These advancements not only lead to cost savings but also open exciting avenues in particle and astroparticle physics research.

\section*{Funding}
\small
\noindent{This work was supported by INFN CSN5 (OREO and MC-INFN projects) and the European Commission through the H2020-INFRAINNOV AIDAINNOVA (G.A. 101004761), H2020-MSCA-RISE N-LIGHT (G.A. 872196) and EIC-PATHFINDER-OPEN TECHNO-CLS (G.A. 101046458) projects. A. Sytov acknowledges support from the H2020-MSCA-IF-Global TRILLION (G.A. 101032975).}


\bibliography{biblio_all.bib}

\end{document}